\documentclass[aps,prc,superscriptaddress,twoside,twocolumn,10pt,floatfix]{revtex4-1}

\usepackage{color}
\usepackage{graphicx}
\usepackage{dcolumn}
\usepackage{bm}
\usepackage{amsmath}
\usepackage{url}
\usepackage{natbib}

\begin{document}


\title{The correlation between the $\alpha$-cluster separation and the neutron S-factor in \(^{12}\text{Be}\) }

\author{Zhilian Cai}
\affiliation{School of Science, Huzhou University, Huzhou 313000, Zhejiang, China}
\author{Qing Zhao} \email[]{Corresponding author}
\email[]{zhaoqing91@zjhu.edu.cn}
\affiliation{School of Science, Huzhou University, Huzhou 313000, Zhejiang, China}
\author{Zaihong Yang}
\affiliation{School of Physics and State Key Laboratory of Nuclear Physics and Technology, Peking University, Beijing 100871, China}
\author{Masaaki Kimura}
\affiliation{Nuclear Reaction Data Centre (JCPRG), Hokkaido University, Sapporo 060-0810, Japan}
\affiliation{Department of Physics, Hokkaido University, Sapporo 060-0810, Japan}
\affiliation{RIKEN Nishina Center, Wako, Saitama 351-0198, Japan}
\author{Bo Zhou}
\affiliation{Key Laboratory of Nuclear Physics and Ion-beam Application (MOE), Institute of Modern Physics, Fudan University, Shanghai 200433, China}
\affiliation{Shanghai Research Center for Theoretical Nuclear Physics, NSFC and Fudan University, Shanghai 200438, China}
\author{Seung-heon Shin}
\affiliation{Department of Physics, Hokkaido University, Sapporo 060-0810, Japan}

\begin{abstract}
The reduced width amplitudes (RWA) and the spectroscopic factor (S-factor) of $\alpha$-cluster and valence neutron in $^{12}$Be are calculated by the generator coordinates method (GCM) with the cluster model. By fixing the distance between the $\alpha$-clusters' generated coordinates, we make a theoretical experiment to analyze the relationship between the $\alpha$-clustering separation and the orbital occupation of the valence neutron in $^{12}$Be. The analysis of the results shows that the percentage of the $\sigma$ orbital occupation in $^{12}$Be is positively related to the clustering separation. 
\end{abstract}

\maketitle


\section{Introduction}
Cluster structure is one of the most interesting phenomena in nuclear physics~\cite{ PDescouvemont_PhysRevC_1989,PR_VONOERTZEN_PhysRep_2006,HISASHI_NPA_1991,Hoyle_AJS_1954,Y.L.Ye_NatRevPhys_2025,K.Wei_NuclSciTech_2024}. The study of clustering in light nuclei is related to various theoretical and experimental aspects, such as cluster resonance states~\cite{Liu_PhysRevLett_2020, Chen_SciBull_2023, Ma_PhysRevC_2021}, molecular states~\cite{P.J.Li_PhysRevLett_2023}, inhomogeneous nuclear matter~\cite{Tanaka_Science_2021, M.Oertel_RevModPhys_2017}, etc. A combined theoretical and experimental study will offer new insights into  the clustering effect on the observed phenomena~\cite{Tanaka_Science_2021}. Thus, finding the direct relationship between the cluster structure and other observables is one of the most important works in theoretical study.

Be isotopes are good candidates for studying the cluster structure since $^{8-12}$Be usually can be treated as two $\alpha$-clusters coupled with several valence neutrons. In the general view from the molecular orbit model, the valence neutrons in these isotopes should occupy the $\pi$-orbit with negative parity. However, it has been known that the last two neutrons in the ground state (0+) of $^{12}$Be dominantly occupy the positive-parity $\sigma$-orbit, which is essentially connected to the breaking of the $N=8$ magic number~\cite{S.D.Pain_PhysRevLett_2006,R.Kanungo_PhysLettB_2010,J.Chen_PhysLettB_2018}. This phenomenon was also attributed to the deformation of the nucleus, which leads to a change in the order of single-particle orbitals and consequently the shell structure~\cite{Kanada_ProgTheorExpPhys_2012,Ito_PhysRevC_2012,A.O.Macchiavelli_PhysRevC_2018}. From the perspective of the cluster model, $^{12}$Be is described as two $\alpha$-clusters and four valence neutrons. As shown by Ito et al. using the generalized two-center cluster model~\cite{Ito_PhysRevC_2012}, molecular orbitals of the valence neutrons are highly sensitive to the configuration of the two alpha clusters. Therefore, the orbital inversion of the valence neutrons should also be closely related to the two-alpha cluster structure. It is thus very interesting to investigate the relationship between the cluster structure and the orbital occupation of the valence neutrons in $^{12}$Be using microscopic theoretical models.

In our previous work~\cite{Zhao_PhysRevC_2022}, we investigate the $\alpha$-cluster structure in $^{10}$Be and $^{12}$Be with the microscopic model calculations. The reduced width amplitude (RWA) and the spectroscopic factor (S-factor) of the $\alpha$-cluster are calculated and analyzed to study the asymptotic behavior and the $\alpha$ formation probability. The RWA is treated as the wave function of a substructure in the parent nucleus, which is a quantitative measure of the cluster structure and can be experimentally probed. The S-factor of the can also be experimentally extracted from the transfer reaction and the proton induced knock-out reaction~\cite{J.Chen_PhysLettB_2018,Wakasa_ProgPartNuclPhys_2017,Z.H.Yang_PhysRevLett_2021,Y.Kubota_PhysRevLett_2021}. Calculating these two quantities provides us a unique access to the relationship between the cluster structure and the valence neutron's orbital occupation.

In this paper, we adopt the generator coordinates method (GCM) combined with the cluster model to calculate the wave functions of $^{12}$Be and $^{11}$Be. After fixing the relative distance of the generated coordinates between the $\alpha$-clusters in their wave functions manually, we then calculate the RWA and S-factor of the valence neutron in $^{12}$Be(g.s.). This is a kind of theoretical experiment that shows us the direct relationship between the clusters separation and the valence neutron's orbital occupation.

This paper is structured as follows. In Section II, the GCM and the calculation formula of RWA and S-factor are briefly introduced. In Section III, we introduce the specific results of this work and the discussion of the results. We make the summary in the last section.

\section{ Theoretical Framework}

\subsection{Hamiltonian and Wave Function}

The Hamiltonian used for the calculation is
\begin{align}
{\cal H}&=\sum_{i}^{A}\hat{t}_{i}-\hat{t}_{cm}+\frac{1}{2}\sum_{ij}^{A}\hat{v}_{NN}(\boldsymbol{r}_{ij})\\
 &+\frac{1}{2}\sum_{ij\in\text{proton}}^{Z}\hat{v}_{C}(\boldsymbol{r}_{ij})+\frac{1}{2}\sum_{ij}^{A}\hat{v}_{ls}(\boldsymbol{r}_{ij})~,\nonumber 
 \label{eq:Ham}
\end{align}
where $\hat{t}_{i}$ and $\hat{t}_{cm}$ represent the kinetic operators of the single-particle and the center of mass, respectively. $\hat{v}_{NN}$, $\hat{v}_{C}$, and $\hat{v}_{ls}$ represent the effective nucleon-nucleon interaction, Coulomb interaction, and spin-orbit interaction, respectively. We use Volkov No.$2$ interaction as the nucleon-nucleon interaction~\cite{Volkov_NucPh_1965},
\begin{align}
\hat{v}_{NN}(\boldsymbol{r}_{ij})=&(W-M\hat{P}^{\sigma}\hat{P}^{\tau}
+B\hat{P}^{\sigma}
-H\hat{P}^{\sigma}) \\
&\times \left[V_{1}\exp{(-\boldsymbol{r}_{ij}^2/c_1^2})+V_{2}\exp{(-\boldsymbol{r}_{ij}^2/c_2^2)}\right]~.\nonumber
\end{align}
, and G3RS potential as the spin-orbit interaction~\cite{Tamagaki_PMNFSD_1968,Nagata_PTPS_1979},
\begin{align}
\hat{v}_{ls}(\boldsymbol{r}_{ij})=V_0^{ls}(e^{-\alpha_1\boldsymbol{r}^2_{ij}}-e^{-\alpha_2\boldsymbol{r}^2_{ij}})\boldsymbol{L}\cdot\boldsymbol{S}\hat{P}_{31}~.
\end{align}
We adopt the interaction parameters as $W=0.42$, $M=0.58$, $B=0.125$, $H=0.125$, and $V_0^{ls} = 2800$ MeV for it can reproduce the $\sigma$-orbital occupation of the valence neutron in the ground state of $^{12}$Be~\cite{Zhao_PhysRevC_2022}. The other parameters are, $V_1=-60.65$ MeV, $V_2=61.14$ MeV, $c_1=1.80$ fm, and $c_2=1.01$ fm for $\hat{v}_{NN}$, $\alpha_1=5.0$ fm$^{-2}$, and $\alpha_2=5.0$ fm$^{-2}$ for $\hat{v}_{ls}$.

We use the generator coordinate method (GCM) combined with the cluster model to construct the wave function of the nucleus. 
The single-particle wave function is expressed as Gaussian form multiplied by the spin and isospin parts $\chi$, $\tau$ as
\begin{align}
\varphi(\boldsymbol{r})=(\frac{2\nu}{\pi})^{3/4}\exp{[-\nu(\boldsymbol{r}-\frac{\boldsymbol{z}}{\sqrt{\nu}})^2+\frac{1}{2}\boldsymbol{z}^2]}\chi\tau\\
\chi=a\chi_{\uparrow}+b\chi_{\downarrow}~;~~\tau=\text{proton or neutron}~.
\end{align}
The Gaussian width parameter is set to be $\nu=1/2b^2$ where $b=1.46$ fm by following the other works\cite{Furumoto_PRC_2018, Itagaki_PRC_2003}. The central of Gaussian is determined by the generate coordinate $\boldsymbol{z}$, which is a complex vector. The real part of generate coordinate $\boldsymbol{R}=\text{Re}(\boldsymbol{z})$ represents the average spatial position of particle, while the imaginary part represents the average momentum.

The $\alpha$-cluster wave function is constructed by the slater determinant with $(0s)^4$ configuration as 
\begin{align}
\Phi_\alpha(\boldsymbol{z}_\alpha)={\cal A}\{\varphi_p(\boldsymbol{z}_\alpha,\uparrow)\varphi_p(\boldsymbol{z}_\alpha,\downarrow)\varphi_n(\boldsymbol{z}_\alpha,\uparrow)\varphi_n(\boldsymbol{z}_\alpha,\downarrow)\}~,
\end{align}
where the spatial coordinates of four particles are set to be the same $\boldsymbol{z}_\alpha$. The basis wave functions for the system are composed of the $\alpha$-cluster and several valence nucleons with the slater determinant as
\begin{align}
\Phi(\boldsymbol{z}_{\alpha,1},\boldsymbol{z}_{\alpha,2}...\boldsymbol{z}_1,\boldsymbol{z}_2...)={\cal A}\{\Phi_{\alpha,1}\Phi_{\alpha,2}...\varphi_1\varphi_2...\}~.
\end{align}

We then perform the angular-momentum projection and the parity projection for each basis wave function to restore the rotational symmetry and the parity.
\begin{align}
\Phi_{MK}^{J\pi}=\frac{2J+1}{8\pi^2}\int d\Omega D_{MK}^{J*}(\Omega)R(\Omega)\hat{P}^{\pi}\Phi.
\end{align}
The total wave function is finally given as the superposition of the basis wave functions
\begin{align}
\Psi_M^{J^\pi} = \sum_{i,K}g_{i}\Phi_{MK,i}^{J^\pi}~.
\end{align}
We can obtain the coefficients $g_i$ and the corresponding eigen-energy by solving the Hill-Wheeler equation.

\subsection{Reduced Width Amplitude (RWA)}
The reduced width amplitude (RWA) is defined as the overlapping integral between the wave functions of the parent nucleus $\Psi$ and the residual nuclei with mass numbers $A_{1}$ and $A_{2}$,
\begin{align}
    Y(a)=a\sqrt{\frac{A!}{(1+\delta_{A_{1}A_{2}})A_1!A_2!}}\left\langle\frac{\delta(r-a)}{r^2}\Psi_{A_{1}}\Psi_{A_{2}}Y_{l}(\hat{r})|\Psi\right\rangle~.
\end{align}
$\Psi_{A_{1}}$ and $\Psi_{A_{2}}$ are the wave functions of two residual nuclei, where $l$ represents the relative angular momentum between them. The RWA can be regarded as the wave function of a substructure in the composed nucleus, whose number of nodes is determined by the quantum number of the residual and the composed nuclei. Thus, it can be direct indicative of valence nucleon orbital occupation in the wave function. This calculation can be done with the Laplace expansion method~\cite{Chiba_ProgTheorExpPhys_2017}. 

Meanwhile, we can calculate the overlap function for the valence nucleon without the single-particle wave function. It is defined with the wave functions of residual and parent nuclei,
\begin{align}
    Y(a)=a\left\langle\delta(r-a)\Psi_{A-1}Y_l(\hat{r})|\hat{a}(\bm{r})|\Psi_A\right\rangle~,
\end{align}
where $\hat{a}(\bm{r})$ is the annihilation operator for one nucleon~\cite{Gaidarov_PhysRevC_1999}. The overlap function has the same physical meaning as the RWA, but the way of counting the number of nodes is different.

The integral of the square of RWA or the overlap function indicates the possibility of finding the substructure in the nucleus, which is called as the spectroscopic factor (S-factor)
\begin{align}
    S=\int_0^\infty Y^2(r) dr~.
\end{align}
In this work, we calculate the RWA for the $\alpha$-cluster in $^{12}$Be to confirm the orbital occupation of its valence nucleons. Meanwhile, we calculate the S-factor for the valence neutron with the overlap function between $^{11}$Be and $^{12}$Be, which reflects the relationship between the $\alpha$-cluster separation and the valence neutron's orbital occupation. We will discuss it in detail in the next section.

\section{Theoretical Results}
\subsection{GCM calculation}

We first calculate the wave functions of the $^{12}$Be, $^{11}$Be, $^{8}$He, and $^{4}$He with the GCM framework as introduced above. These wave functions will be used for the further calculations in this work. The energy results and corresponding Q-values are shown in Table~\ref{table:ene},
\begin{table}[htbp]
  \begin{center}
    \caption{The binding energies and the corresponding Q-values of $^{12}$Be, $^{11}$Be, $^{8}$He, and $^{4}$He. ``Expt." indicates the experimental data, which is referred from NNDC database~\cite{NNDC}, ``Cal." indicates the calculated results. ``Q-v(Exp/Cal)" indicate the experimental/calculated Q-values from $^{12}$Be to $^{11}$Be and $^{8}$He + $^{4}$He, respectively. All the units are in MeV. \label{table:ene}}
    \vspace{2mm}
 \begin{tabular*}{8cm}{ @{\extracolsep{\fill}} l c c c c}
    \hline
                      &Expt.    &Q-v(Exp)&Cal.     &Q-v(Cal)\\
    \hline
$^{12}$Be($0^+$)      &$-68.65$ &        &$-71.14$ &       \\
$^{11}$Be($1/2^+$)    &$-65.48$ &$-3.17$ &$-70.51$ &$-0.63$\\
$^{4}$He($0^+$)       &$-28.30$ &        &$-27.57$ &       \\
$^{8}$He($0^+$)       &$-31.40$ &$-8.95$ &$-34.66$ &$-8.91$\\
   \hline
  \end{tabular*}
  \end{center}
\end{table} 
where we can see that the binding energies are in good agreement with the experimental data. More importantly, the order of $^{12}$Be and $^{11}$Be is correct. Since we aim to deal with the virtual wave functions in further calculations, the absolute value of the binding energy is not as important as in other theoretical works, but the order of the states and the orbital occupation of the valence neutron are more critical.

\begin{figure}
    \centering
    \includegraphics[width=1.0\linewidth]{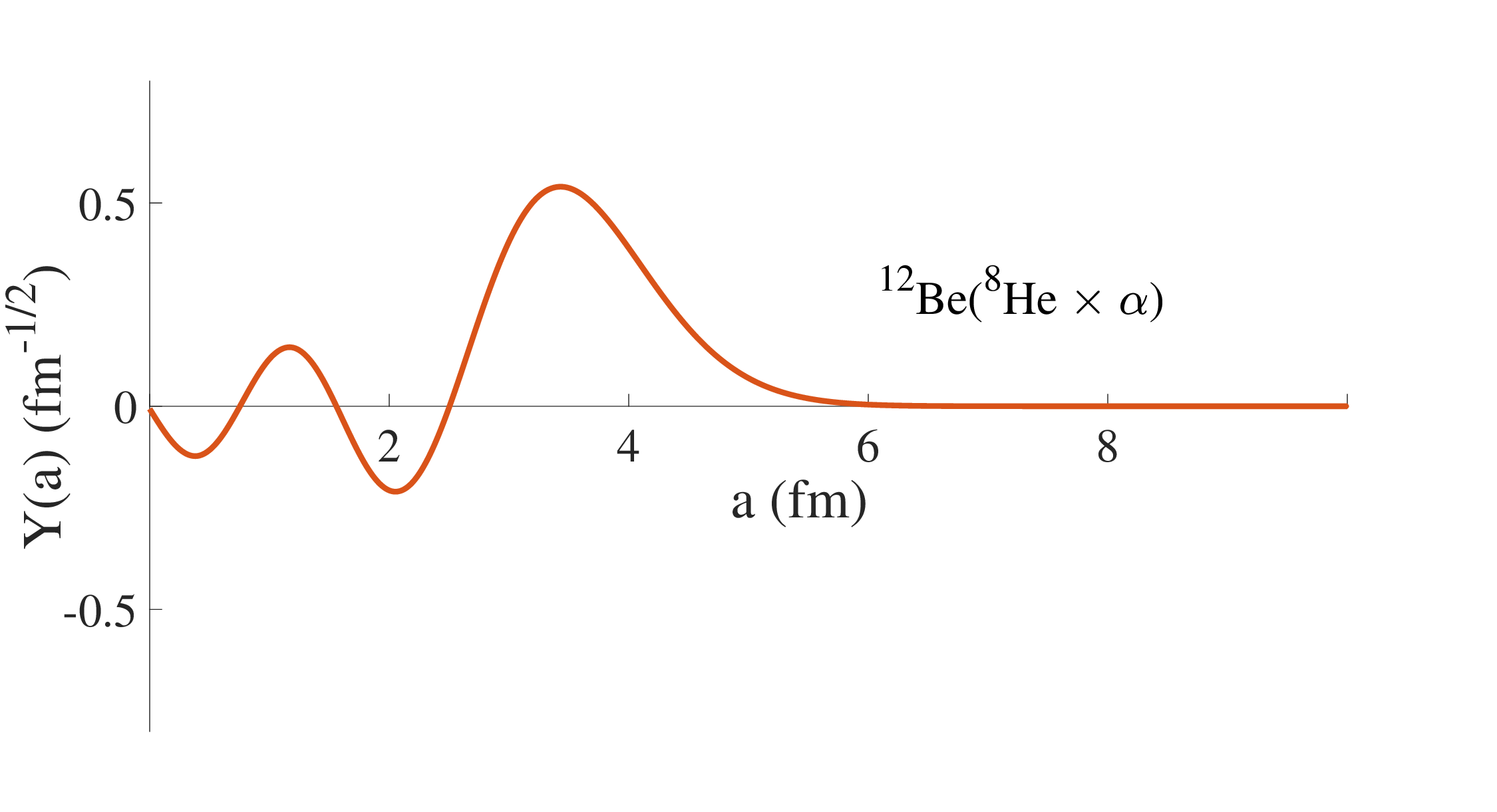}
    \caption{The calculated $\alpha$-RWA of $^{12}$Be to $^{8}$He + $^{4}$He channel. All the nuclei are in their ground state.}
    \label{fig:arwa}
\end{figure}

The $\sigma$-orbital occupation in the ground state of $^{12}$Be is known as one of the examples of the breaking of the magic number $N=8$. As introduced in our previous work, the orbital occupation of the valence neutron in the calculated wave function of $^{12}$Be can be clearly confirmed from its $\alpha$-RWA result. The result is shown in Fig.~\ref{fig:arwa}.
It can be seen that there are three nodes in the RWA, which indicates that the last two neutrons in the calculated wave function of $^{12}$Be are occupying $\sigma$-obit as in reality. Next, we can continue our further investigations based on the current wave functions.

\subsection{Correlation between $\alpha$-cluster separation and neutron S-factor}

\begin{figure}
    \centering
    \includegraphics[width=1.0\linewidth]{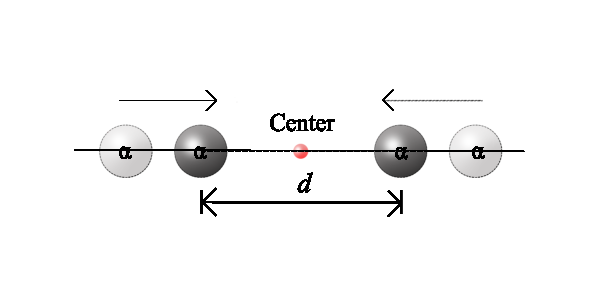}
    \caption{The diagram of illustrating the specific operation of modifying the distance between two clusters. The solid sphere represents the $\alpha$ cluster in the initial basis wave function. The translucent sphere represents the location of the modified clusters. In this work, we adjusted the distance of clusters in $^{11}$Be and $^{12}$Be to $d=1/1.5/2/2.5/3$ fm.}
    \label{fig:changer}
\end{figure}

In this work, we define the $\alpha$-cluster separation with the spatial generate coordinates of clusters as $|\boldsymbol{R}_{\alpha,1}-\boldsymbol{R}_{\alpha,2}|$, which represents the average distance of clusters. Since the GCM wave function is a superposition of many basis wave functions with different generated coordinates for the clusters, we modify and fix their coordinates to make their distance a constant. The way of fixing the distance is a manual work for each basis wave function. First, we find out the center of mass between two clusters $\boldsymbol{R}_c = \frac{1}{2}(\boldsymbol{R}_{\alpha,1}+\boldsymbol{R}_{\alpha,2})$. Second, we modify two clusters' spatial coordinates to $\boldsymbol{R}^\prime_i$ to make their relative distance become $d$ while keeping their center of mass and the relative direction unchanged. The relationship between $\boldsymbol{R}^\prime_{\alpha,i}$ and $\boldsymbol{R}_{\alpha,i}$ is calculated as below.
\begin{align}
\boldsymbol{R}^\prime_{\alpha,i} = d\frac{(\boldsymbol{R}_{\alpha,i}-\boldsymbol{R}_c)}{|\boldsymbol{R}_{\alpha,1}-\boldsymbol{R}_{\alpha,2}|}+\boldsymbol{R}_c.
\end{align}
Finally, we calculate the total wave function with these set of modified basis wave functions with GCM. A diagram of how to fix the distance between two clusters is shown in Fig.~\ref{fig:changer}.
In this work, we fix the cluster distance for $^{11}$Be and $^{12}$Be simultaneously. Therefore, the changing of the overlap function between them only reflects the feature of the valence neutron. As the comparison, we adjust the cluster distance as $d = 1/1.5/2/2.5/3$ fm.

\begin{figure}
    \centering
    \includegraphics[width=1.0\linewidth]{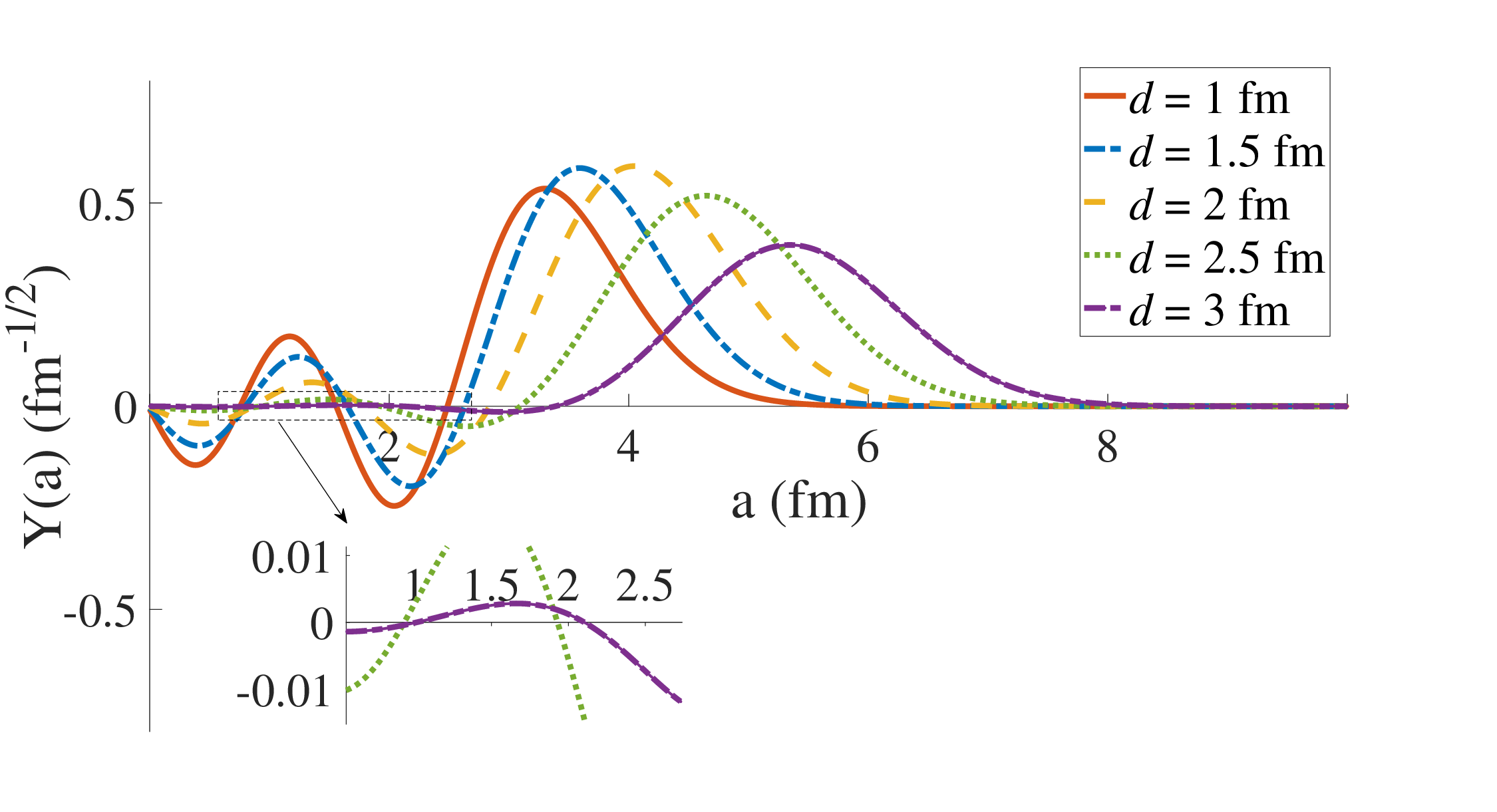}
    \caption{The $\alpha$-RWAs of the ground state of $^{12}$Be which has different clusters distance. The sub-figure shows the local enlargement.}
    \label{fig:ARWA}
\end{figure}

We still need to confirm that the orbital structure of the valence neutron is not fundamentally altered due to the changes in the generated coordinates of the two clusters. We calculate the $\alpha$-RWAs for each fixed distance and show the results in Fig.~\ref{fig:ARWA}, 
where we can see the same number of nodes as the RWA with the original $^{12}$Be wave function. It indicates the unchanged orbital occupation of the valence neutron in these virtual wave functions. In addition to this, we can also see that the RWAs are shifted to the right along with the increase in the cluster distance, which follows the diagram of the $\alpha$-clusters in $^{12}$Be. The valence neutron in $^{11}$Be($1/2^+$) is must be in the $\sigma$-orbit because of the spin-orbit coupling, therefore we do not need to check it with the $\alpha$-RWA.

\begin{figure}
    \centering
    \includegraphics[width=1.0\linewidth]{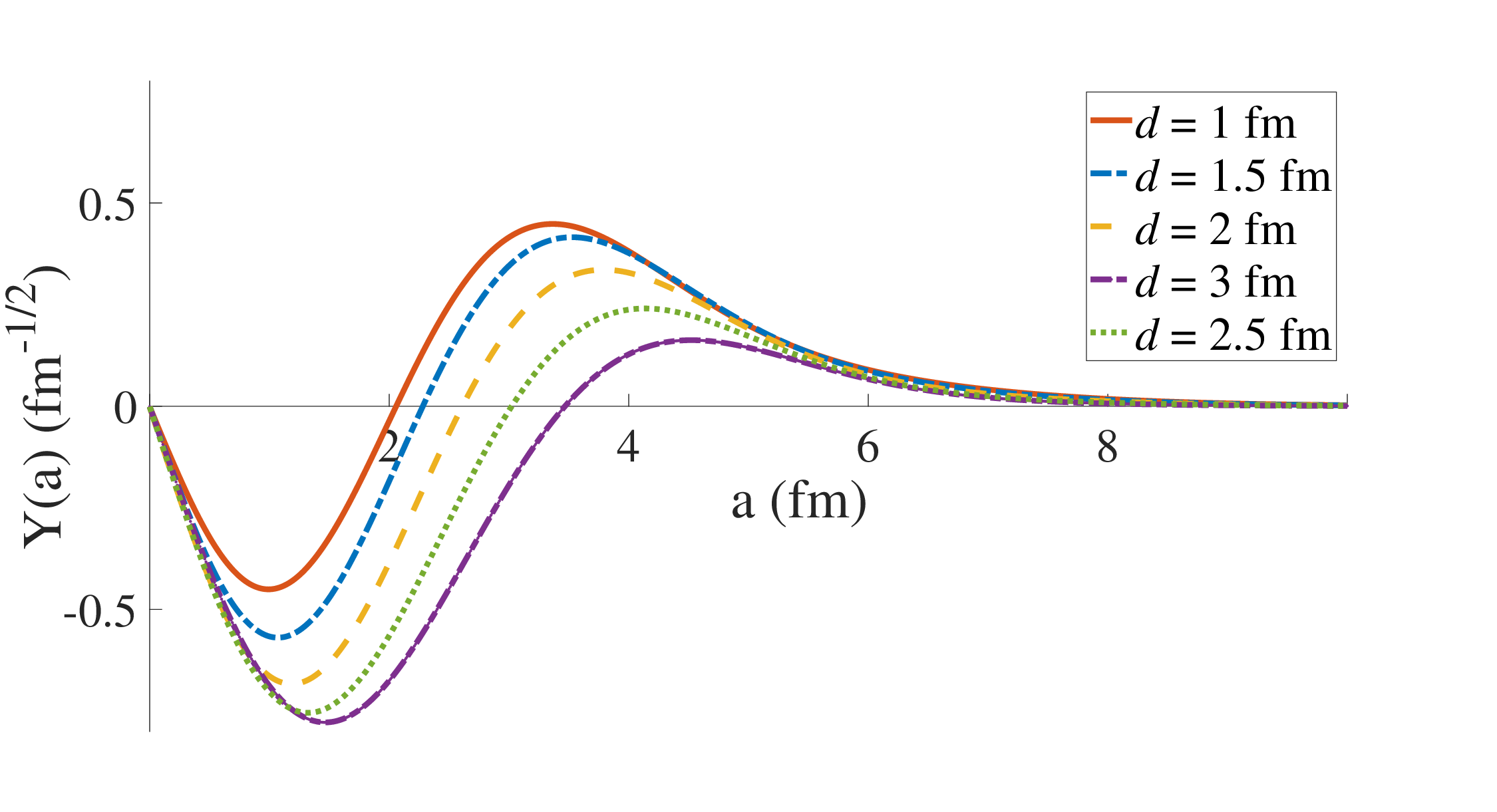}
    \caption{The overlap function results between $^{11}$Be and $^{12}$Be at different $\alpha$ clustering distances. The $^{11}$Be and $^{12}$Be are all in their ground state from the calculation.}
    \label{fig:NRWA}
\end{figure}

\begin{figure}
    \centering
    \includegraphics[width=1\linewidth]{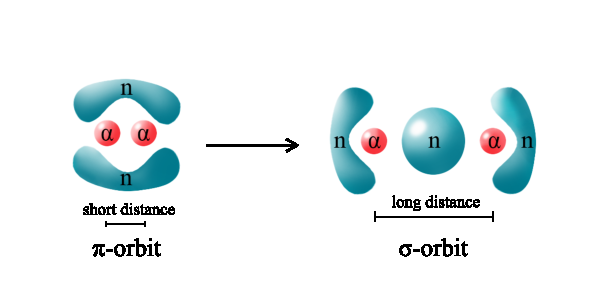}
    \caption{The diagram of the nuclear density for $\pi$-orbit and $\sigma$-orbit.}
    \label{fig:model}
\end{figure}

After fixing the generated coordinates' distance between the $\alpha$-clusters, we calculate the overlap function between the wave functions of $^{11}$Be and $^{12}$Be. The overlap function represents the wave function of a single neutron in $^{12}$Be, which has a similar meaning as the RWA. Especially, because the cluster structure in these two nuclei are fixed to be the same, the overlap results only reflect the behavior of the valence neutron effected with difference cluster separation. The results with different $\alpha$-cluster distances are shown in Fig.~\ref{fig:NRWA}.
From this figure, we can notice that the overlap function of the single-neutron is directed with the $\alpha$-clustering separation. Along with the increasing distance, the inner part of the single-neutron overlap function amplitude is enhanced and the outer part is suppressed. It follows the pictures of the $\pi$- and $\sigma$-orbit's structures, that the $\pi$-orbit has one node and moves around the system, while the $\sigma$-orbit has two nodes so that has a distribution at the inner part. In the nucleus, the position of the clusters relates to the position of the nodes of the single-particle orbitals. The larger separation between two clusters in $^{12}$Be may gives more space to form the $\sigma$-orbit for the valence neutron. A diagram of the above analysis is shown in Fig.~\ref{fig:model}.

\begin{figure}
    \centering
    \includegraphics[width=1.0\linewidth]{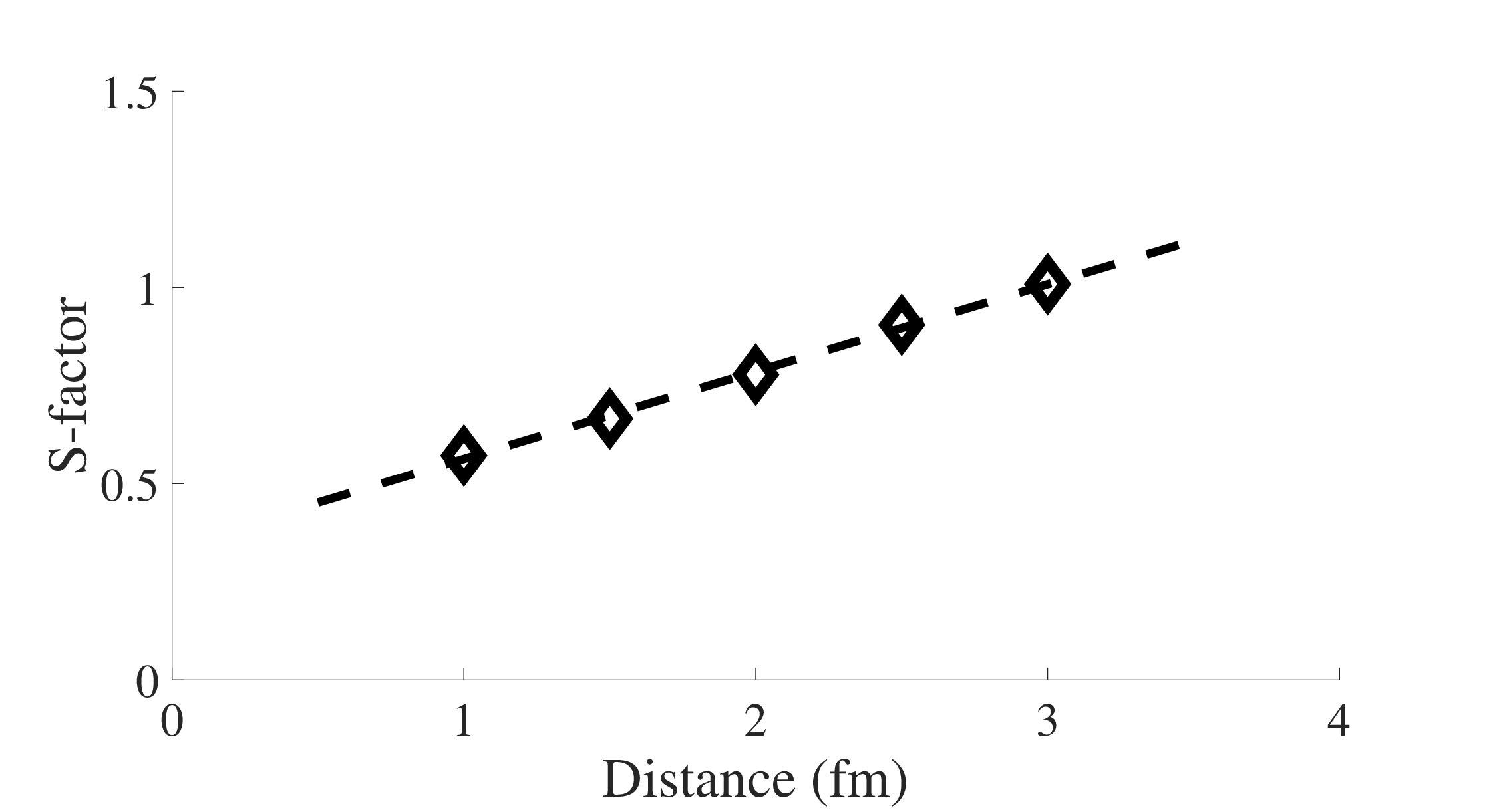}
    \caption{The S-factors of the single-neutron in $^{12}$Be(g.s.) with different fixed clusters distance. The dashed line is the straight fitting line for the results.}
    \label{fig:sfactor}
\end{figure}

Since the last neutron in $^{11}$Be has $100\%$ to occupy the $\sigma$-orbit, the S-factor between $^{11}$Be and $^{12}$Be with the same cluster separation directly indicates the percentage of the valence neutron in $^{12}$Be occupying the $\sigma$-orbit. We show the S-factor results along with the increasing of the distance in Fig.~\ref{fig:sfactor}.
In this figure, the S-factor shows a positive correlation with the $\alpha$-cluster separation, which is again consistent with our picture of the orbits. What is even more interesting is that we can clearly see a linear relationship between the S-factor and the distance of clusters' generated coordinates. It suggests some simpler relationship between orbital occupation of valence neutron and cluster separation. This result may provide a new perspective to the experimental study of the nuclear orbital occupation and the cluster structure.

\section{Summary}
We first calculate the wave functions of the $^{12}$Be, $^{11}$Be, $^{8}$He, and $^{4}$He with the GCM framework and confirm the valence neutrons in the ground state of $^{12}$Be dominantly occupy the $\sigma$-orbit, in accordance with the well-known breaking of the N=8 shell closure. By fixing the spatial distance between two $\alpha$-clusters' generated coordinates in the wave functions of $^{12}$Be and $^{11}$Be manually, we calculate the corresponding overlap function and the S-factor of the valence neutrons in $^{12}$Be. The results corroborate the conjecture that the evolution of the neutron molecular orbitals is closely tied to the configuration of the two alpha clusters---namely, the increasing separation between two $\alpha$-clusters can enhance the $\sigma$-orbit occupation. Furthermore, our result reveals a linear positive correlation between the $\sigma$-orbital occupation and the $\alpha$-clusters separation. Notably, both the two-$\alpha$-cluster structure and the neutron S-factor can be measured experimentally by using the proton-induced knockout reactions~\cite{P.J.Li_PhysRevLett_2023, Z.H.Yang_PhysRevLett_2021}. This intriguing correlation may thus be probed by incorporating the predicted RWA of $\alpha$-clusters and the overlap function of the valence neutrons from our theoretical calculations into the reaction theories.

While the linear correlation between the S-factor and the alpha-cluster separation appears promising, further work is needed to refine our theoretical calculations. One notable limitation lies in the reliance on the virtual wave functions of $^{12}$Be and $^{11}$Be in the current study, which inevitably deviates from reality. Additionally, there is some ambiguity in the definition of cluster separation. In this work, the $\alpha$-$\alpha$ separation is determined using the generated coordinates of the clusters. However, these coordinates do not represent the physical positions of the clusters but rather the centers of their Gaussian wave packets. More accurately, they should be interpreted as the intrinsic average positions of the clusters, given the particle exchange between the clusters and the valence neutrons, which prevents localization of the clusters.

\begin{acknowledgments}
This work was supported by the National Key R\&D Program of China (No. 2023YFE0101500), the National Natural Science Foundation of China [Grant Nos. 12305123, 12175042, 12275081, 12275082, 12275006], and JSPS KAKENHI [Grant Nos. 19K03859, 21H00113 and 22H01214]. Numerical calculations were performed in the Cluster-Computing Center of School of Science (C3S2) at Huzhou University.
\end{acknowledgments}

\end{document}